\documentclass[utf8]{frontiersSCNS} % for Science, Engineering and Humanities and Social Sciences articles

\setcitestyle{round} % for Physics and Applied Mathematics and Statistics articles

\usepackage{url,hyperref,lineno,microtype}
\definecolor{bleudefrance}{rgb}{0.19, 0.55, 0.91}
\hypersetup{colorlinks=true, citecolor=bleudefrance, linkcolor=bleudefrance, urlcolor=bleudefrance}
\usepackage[onehalfspacing]{setspace}
\usepackage{threeparttable} 
\usepackage{sidecap}

\def\keyFont{\fontsize{8}{11}\helveticabold }
\def\firstAuthorLast{Fern\'andez-Valenzuela} %use et al only if is more than 1 author
\def\Authors{Estela Fern\'andez-Valenzuela\,$^{1,*}$}
\def\Address{$^{1}$Florida Space Institute, University of Central Florida, Orlando , FL, USA}
\def\corrAuthor{12354 Research Parkway\\
Partnership 1 Building, suite 211\\
Orlando, FL 32826-0650, USA}

\def\corrEmail{estela@ucf.edu}

\begin{document}
\onecolumn
\firstpage{1}

\title[Modeling long-term photometric data of TNOs and Centaurs]{Modeling long-term photometric data of trans-Neptunian objects and Centaurs} 

\author[\firstAuthorLast ]{\Authors} %This field will be automatically populated
\address{} %This field will be automatically populated
\correspondance{} %This field will be automatically populated

\extraAuth{}% If there are more than 1 corresponding author, comment this line and uncomment the next one.
%\extraAuth{corresponding Author2 \\ Laboratory X2, Institute X2, Department X2, Organization X2, Street X2, City X2 , State XX2 (only USA, Canada and Australia), Zip Code2, X2 Country X2, email2@uni2.edu}

\maketitle

\begin{abstract}

\noindent
Trans-Neptunian objects and Centaurs are small solar system bodies that reside in the outer parts of the solar system. These objects present photometric behaviours that are influenced due to a change in their aspect angle. Using absolute photometric measurements and rotational light-curves at different location on their orbits allow to model their photometric behaviour and obtain physical properties such as pole orientation, shape, density, and even detecting different peculiarities as departure from hydrostatic equilibrium or rings. In this work, I present how these models are performed in order to extract different physical properties. A review of the objects for which their long-term photometric behaviour has been modeled is also given.

%%% Leave the Abstract empty if your article does not require one, please see the Summary Table for full details.
\section{}

\tiny
 \keyFont{ \section{Keywords:} TNOs, Centaurs, imaging, long-term variability} %All article types: you may provide up to 8 keywords; at least 5 are mandatory.
\end{abstract}

\section{Introduction}

Trans-Neptunian Objects (TNOs) are small solar system bodies that orbit the Sun with orbital semi-major axes larger than that of Neptune but smaller than where the Oort Cloud begins \citep{Oort1950,Gladman2008}. They were formed outside of the frost line, namely, far enough from the Sun so that chemical compounds with low sublimation points (i.e., ices such as N$_2$, CO, CH$_4$) were able to survive and condensate to form these objects. TNOs are therefore composed of a mixture of ices and silicates incorporated as grains during accretion from the solar nebula. Due to their large distances that separate them from the Sun, TNOs have undergone few thermo-chemical processes when compared to other small bodies in the solar system. Hence, in some sense, these objects are time-capsules containing almost pristine material from when the solar system was formed. Thus, the study of TNOs provides important information about the primitive solar nebula. 

When studying TNOs, it is also common to include the so-called Centaurs because these objects originally formed and have spent most of their life-time in the trans-Neptunian region \citep{Horner2004a}, but have been injected into inner parts of the solar system due to planetary encounters, mostly with Neptune \citep{Fernandez1980}. Therefore, Centaurs present an excellent opportunity to study smaller trans-Neptunian objects much closer to the Earth, providing a better characterization of their physical properties. 

\subsection{Synopsis on TNOs formation}
\label{sec:collisional_evolution}

Current theoretical models suggest that the formation of planetesimals in the trans-Neptunian region was driven through the streaming instability, achieving up to $\sim100$ km in size \citep{Johansen2015,Simon2016}. These planetesimals would initially have a phase of collisional coagulation, with the largest objects acquiring a size of 300 – 400 km in radius \citep{Lambrechts2016}, and then those bodies would accrete individual pebbles \citep{Lambrechts2014,Lambrechts2016}. These physical processes are able to reproduce the primordial size distribution in the trans-Neptunian disk \citep{Morbidelli2020}. Then, the massive disk phase took place, during which collisional activity was very intense \citep{Morbidelli2015}, and TNOs suffered strong impacts that, without complete disruption, would convert them into gravitational aggregates. This implies that TNOs $>100$ km can be treated as a fluid in a first approximation. 

\subsection{Equilibrium shapes for homogeneous bodies}

\begin{figure}[h!]
\begin{center}
\includegraphics[width=\textwidth]{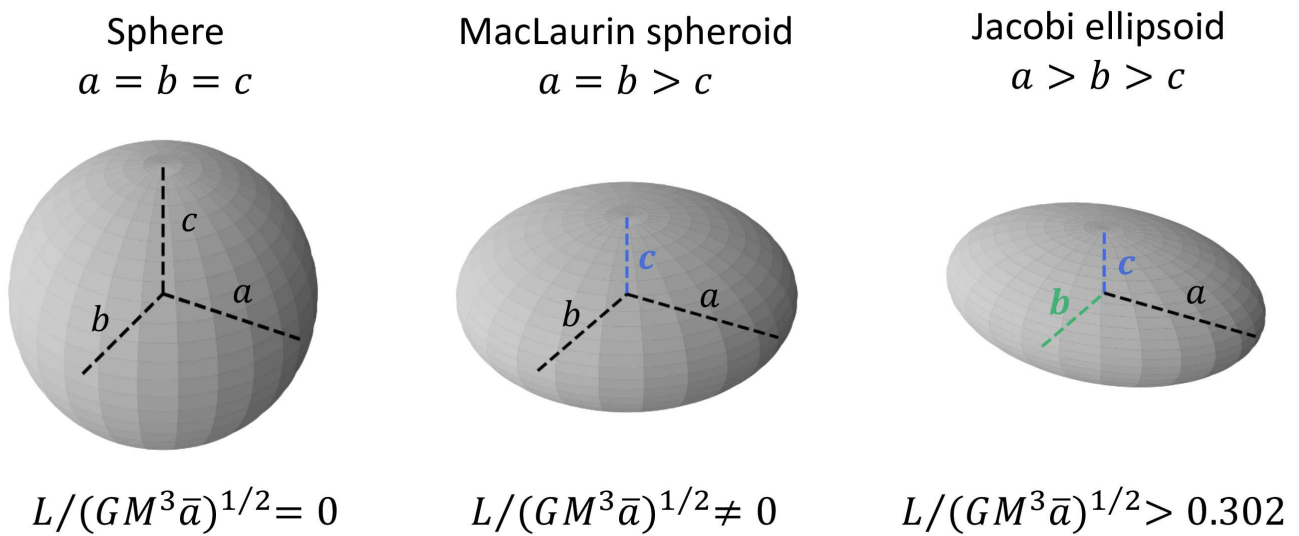}
\end{center}
\caption{Equilibrium shapes for homogeneous bodies.}\label{fig:shapes}
\end{figure}

A fluid, under the assumption of hydrostatic equilibrium with null angular momentum, $L$, would adopt a spherical shape with semi-axes $a$, $b$, and $c$ (where $a=b=c$, see left panel in figure \ref{fig:shapes}). If the fluid rotates, i.e., $L\ne0$, the fluid would adopt a MacLaurin spheroid shape (with $a=b>c$). This happens until $L=0.302\times(GM^3\overline{a})^{1/2}$, where $G$ is the gravitational constant, $M$ is the mass of the fluid and $\overline{a}$ is the radius of a sphere of the same mass as the spheroid or ellipsoid that is given by $(a^2c)^{1/3}$, or $(abc)^{1/3}$, respectively. Beyond this point, the fluid can adopt a Jacobi ellipsoid shape (with $a>b>c$), as described with the mathematical formalism of \cite{Chandrasekhar1969}.

\section{Short-term variability of TNOs and Centaurs}

In order to understand the long-term photometric evolution of TNOs and Centaurs, first, we need to understand short-term variability and the reasons behind the different behaviors. The short-term variability of TNOs and Centaurs (and small bodies in general) is generally produced by the body's rotational modulation, and it is commonly recognized as its rotational light-curve. It is also possible that the short-term variability might be caused by a binary system when one of the components is being occulted by the other, however, this would not be a rotational light-curve, because the variability would not be due to the rotational modulation of the object. Note that these light curves have specific shapes that differ from those explained below and should be carefully analyzed as explained in \cite{Harris2020}. The rotational light-curve of a small body is generally due to albedo variability on the surface, a non-spherical body shape, or a combination of both. Specifically, for TNOs and Centaurs, due to their large sizes, tumbling effects are not expected.

For MacLaurin spheroids, the rotational light-curve is expected to be featureless or flat, i.e., with no variability, unless the object presents an albedo spot (or several) that produces a variation in the amount of reflected light by the surface. This is the case of the dwarf planet Pluto, with a MacLaurin spheroid shape that presents a single-peaked rotational light-curve induced by the large N$_2$ albedo spot on its surface, although this case is somewhat atypical since the variability produced by the albedo spot is of 0.26 mag in V-band \citep{Buratti2003}. We do not expect a variability larger than 0.2 mag to be due to an albedo spot in asteroids \citep{Degewij1979,Magnusson1991}. This value is even smaller for TNOs, which usually do not produce variabilities larger than 0.1 mag due to albedo spots \citep{Thirouin2010}. However, this is for objects much smaller than Pluto that are not able to retain volatile material during large periods of time (i.e., over the age of the solar system). Therefore, when the rotational light-curve presents a large peak-to-valley amplitude (above 0.2 mag), it is common to think that the variability is due to the body shape, with the rotational light-curve presenting a double-peaked shape. 

An inspection of the rotational light-curve is important to determine whether the shape is double-peaked or single-peaked. For instance, a double-peaked rotational light-curve presenting different depths among the minima and maxima might indicate that a tri-axial shape is more likely than an oblate spheroid with an albedo spot. This is the case of (20000) Varuna \citep[e.g.,][]{Jewitt2001,Fernandez-Valenzuela2019}, (84922) 2003 VS$_2$ \citep{Vara-Lubiano2020}, and (470599) 2008~OG$_{19}$
\citep{Fernandez-Valenzuela2019}, among others. Note that detecting the difference between the brightness minima and/or the brightness maxima of the rotational light-curve requires high quality data, as is the case in the mentioned examples. In an opposite scenario, we found Pluto. If we plot Pluto's rotational light-curve doubling its rotation period, the resulting rotational light-curve will present a double-peaked shape with two identical minima and maxima. It is very unlikely that a solar system object with a tri-axial shape would be identical along with its rotational modulation, being this an indication of an object with a single-peaked rotational light-curve due to an albedo spot.

For objects whose rotational light-curve is due to an albedo spot, i.e., a single-peaked rotational light-curve, the amplitude will indicate the difference in brightness produced by the different materials on their surfaces. On the other hand, for objects whose rotational light-curve is due to body shape, i.e., the amplitude of their rotational light-curve depends on the axial ratios and the aspect angle of the body\footnote{The aspect angle of a solar system body is defined as the angle between the line of sight and its rotational axis} as follows:
\begin{equation}
    \label{eq:amplitude_delta}
    \Delta m = -2.5 \log\left[ \frac{b}{a} \left( \frac{a^2\cos^2(\delta) + c^2\sin^2(\delta)}{b^2\cos^2(\delta) + c^2\sin^2(\delta)} \right)^{1/2}\right]
\end{equation}
\cite[e.g.,][]{Surdej1978,Pospieszalska-Surdej1985} where $\Delta m$ is the peak-to-valley amplitude of the rotational light-curve, $\delta$ is the aspect angle, and $a$, $b$ and $c$ are the semi-axes of the tri-axial body (see figure \ref{fig:shapes}). If we consider that the object has an aspect angle of $90^{\circ}$, the equation will be simplified as:
\begin{equation}
    \Delta m = -2.5\log\left(\frac{b}{a}\right).
\end{equation}

However, a solar system object might present any aspect angle from 0 to $180^\circ$, and therefore, from a single rotational light-curve it is only possible to obtain an upper limit of the axes ratio $b/a$.

\section{Long-term variability of TNOs and Centaurs}
\label{sec:long-term}

The long-term variability of a TNO or Centaur is generally due to a change in the aspect angle of the body, $\delta$, which is given by the equation:
\begin{equation}
\label{eq:aspect_angle}
\delta = \frac{\pi}{2} - \sin\left[\sin(\beta_{\rm e})\sin(\beta_{\rm p}) + \cos(\beta_{\rm e})\cos(\beta_{\rm p})\cos(\lambda_{\rm e}-\lambda_{\rm p})\right],
\end{equation}
where $\beta_{\rm e}$ and $\lambda_{\rm e}$ are the ecliptic latitude and longitude of the object reference frame and $\beta_{\rm p}$ and $\lambda_{\rm p}$ are the ecliptic latitude and longitude of the pole orientation of the body, i.e., the coordinates indicating the orientation (not direction\footnote{Because TNOs and Centaurs are far from the Earth, it is only possible to observe them at small phase angles (with a maximum of around $2^{\circ}$ and $5^{\circ}$ for TNOs and Centaurs, respectively), and therefore, the techniques used to obtain the direction of the rotational axis in asteroids can not be carried out for objects beyond Jupiter.}) of the rotational axis. 
Therefore, observing rotational light-curves at different orbit's locations will result in different values of $\Delta m$, allowing the obtainment of the pole orientation of the body, by fitting equations \eqref{eq:amplitude_delta} and \eqref{eq:aspect_angle} to the observational data. This method has been extensively use for asteroids \citep{Magnusson1989}. However, for outer solar system objects, this method was applied for the first time by \cite{Tegler2005} to the Centaur (5145) Pholus. They used observations of Pholus in three different epochs that produced three different values of Pholus' rotational light-curve amplitude \cite[0.15, 0.39, and 0.60 mag published in][respectively]{Buie1992,Farnham2001,Tegler2005}, obtaining two different pole orientations (with their supplementary direction also possible) and the axial ratios of Pholus tri-axial shape.

The change in the aspect angle does not only affect $\Delta m$, but also the absolute magnitude, $H$\footnote{The absolute magnitude of a solar system body is the apparent magnitude of the object when located at 1 au from the Sun, 1 au from the Earth and 0$^\circ$ aspect angle.}, since the amount of area that is reflecting light from the Sun of a tri-axial body varies with $\delta$. For a given epoch (i.e., a specific $\delta$), the area of the body will be the average value from the rotational modulation of the object, i.e.,
\begin{equation}
A(\delta) = \frac{{\rm Area_{max}(\delta)} + {\rm Area_{min}(\delta)}}{2},
\end{equation}
where the maximum and minimum area exposed are given by the following equations:
\begin{equation}
{\rm Area_{max}} = \pi\,a[b^2\cos^2(\delta)+c^2\sin^2(\delta)]^{1/2},
\end{equation}
and
\begin{equation}
{\rm Area_{min}} = \pi\,b[a^2\cos^2(\delta)+c^2\sin^2(\delta)]^{1/2}.
\end{equation}
Thus, the absolute magnitude is given by the equation:
\begin{equation}
\label{eq:absolute_mag}
    H=-M_{\odot}+2.5\log\left(\frac{\pi C^2}{pA(\delta)}\right),
\end{equation}
where $M_\odot$ is the absolute magnitude of the Sun, $C=1330\pm18$ km is a constant \citep{Masiero2021}, and $p$ is the geometric albedo of the object.

The long-term variability of the absolute magnitude, in which a change of the aspect angle is considered\footnote{We should not confuse this long-term variability, that is driven by the change in the aspect angle, with changes of brightness due to outburst or sudden cometary activity. Although some things can be done in this regard, as we describe for the case of Chiron, this kind of activity involves increase and decrease of brightness that behave completely different from the ones we are describing here.}, was first modeled for the centaur (10199) Chariklo in \cite{Duffard2014}, where a compilation of $H$ values from the literature was done. \cite{Duffard2014} found variability of 0.6 mag in a time span of 10 years. One might think that this $H$ magnitude variability could be due to the rotational modulation of the object, if those measurements were not corrected from it; however, this variability is much larger than any of Chariklo's rotational light-curve amplitudes found in the literature \citep[e.g.,][]{Peixinho2001,Fornasier2014}. Therefore, this change in magnitude could not be due to rotational variability but due to a change in the aspect angle. From the stellar occultation produced by this object in 2013, \cite{Braga-Ribas2014} obtained the tri-axial shape of the object (i.e., values for $a$, $b$ and $c$) and two possible solutions for the pole orientation (with their supplementary directions also possible). Using those results, \cite{Duffard2014} found that only one of the solutions of the pole orientation is compatible with the observational data.
The case of Chariklo is somewhat special since it possesses a ring system, which contributes enormously to this variation in absolute magnitude. This is because the variation of the exposed area of a ring is much larger than for a tri-axial body (see panel c in figure \ref{fig:Chariklo_examples} and section \ref{sec:peculirities}). In this case, equation \ref{eq:absolute_mag} needs to be modified as follows:
\begin{equation}
    H=-M_{\odot}+2.5\log\left(\frac{\pi C^2}{pA(\delta)+p_{\rm R}A_{\rm R}(\delta)}\right),
\end{equation}

where $p_{\rm R}$ is the ring's albedo and $A_{\rm R}=\pi R^2|\cos(\delta)|$ is the area of the ring as a function of the aspect angle. Currently, there are four Centaurs and one TNO for which their long-term photometric behavior has been modeled: Pholus \citep{Tegler2005}, Chariklo \citep{Duffard2014}, (2060) Chiron \citep{Ortiz2015}, (54598) Bienor \citep{Fernandez-Valenzuela2017}, and Varuna \citep{Fernandez-Valenzuela2019}. For those objects, shape, pole orientation and, more importantly, density have been obtained, with the exception of Chiron, for which its past comae behavior prevented from obtaining a good constrain of its shape, and, therefore, only allowing an estimation of the density). 

Density is one of the most important properties to understand the interiors of the small bodies in the solar system, and obtaining statistically meaningful results is pivotal for comprehending the solar system formation and evolution. However, measuring density in small objects on the outer system is challenging. Density is usually obtained by studying binary systems or objects hosting satellites, but at the distances at which TNOs are found, the separation is generally too small to resolve the system and obtain its orbital properties. The Hubble Space Telescope (HST) was very successful when the High Resolution Camera was available. Currently, the Wide Field Camera on board HST can resolve objects separated over 0.04'', which is very limited for distances over 30 au. The James Webb Space Telescope (JWST) might provide a better resolution with NIRCam thanks to the exquisite Point Spread Function, but this needs still to be evaluated once the JWST is working. In any case, this technique is biased towards finding satellites or binary systems separated by at least the resolution of the instrument from which they are observed.  

For single objects, studying the shape and rotation state is the only way to measure their density from ground-based observations. Good estimations can be obtained through the study of rotational light-curves \cite[see, for instances,][]{Tancredi2008,Rambaux2017}. However, the combination of multi-chord stellar occultations with rotational light-curves would provide the most accurate results. Rotational light-curves are needed not only to measure the rotation period but also to obtain the rotational phase at which the stellar occultation occurs, thus providing the three-dimensional shape of the object. However, the statistic suggests that in order to observe a multi-chord stellar occultation, it is required to have at least 15 different observing sites distributed along the shadow path, which is not always possible \citep{Ortiz2020}.

An alternative technique would be the study of the long-term photometric behavior, fitting observational data and obtaining not only the pole orientation but also the shape (i.e., axial ratios). Due to the distances at which TNOs and Centaurs orbit the Sun, this technique has the inconvenience of requiring measurements taken with a time span in the order of decades. This is, the arc of the orbit traveled by the object needs to be large enough to produce a significant change in its aspect angle. Currently, it has been almost 30 years since the first TNO, 1992 QB$_1$ (without considering Pluto), was discovered \citep{Jewitt1993} and the community has accumulated enough data of some objects that this might be a solution to obtain statistically meaningful results for the outer solar system. This technique would also require us to assume hydrostatic equilibrium in order to obtain the density of these objects, however, as we discuss in section \ref{sec:collisional_evolution}, this seems to be a reasonable solution.

\section{Extracting peculiarities from long-term variability}
\label{sec:peculirities}

\begin{figure}[h!]
\begin{center}
\includegraphics[width=\textwidth]{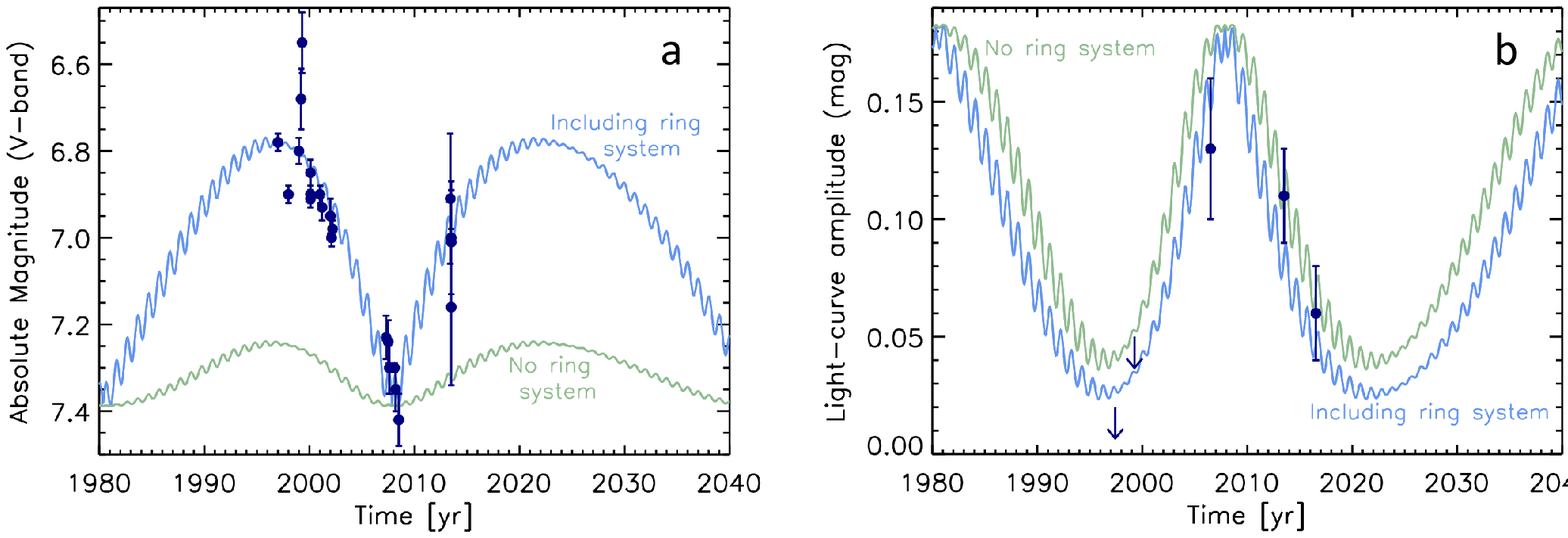} % This is a *.eps file
\end{center}
\caption{Comparison of Chariklo's models for the absolute magnitude (a) and the amplitude (b), with and without the ring system. Dark blue circles and arrows (upper limits) represent observational data from the literature, see \cite{Duffard2014}, and references therein for panel (a); and \cite{Davies1998,Peixinho2001,Galiazzo2016,Fornasier2014,Leiva2017} for panel (b). Panel (c) provides the aspect angle of Chariklo over one orbital period.}\label{fig:Chariklo_examples}
\end{figure}

The long-term photometric data can serve as indications of different peculiarities, such as rings, in small bodies. Rings are located in the equator of the body, presenting very similar pole orientations (i.e., errors related to assuming the same pole orientation for the host body and the rings are negligible for our purpose). Both absolute magnitude and amplitude of the rotational light-curve will be affected if the body presents a ring (or rings). In the case of the absolute magnitude, when the system is found in an edge-on orientation, i.e., the aspect angle is close to $90^{\circ}$, the area reflecting light from the ring is negligible, with almost all the light being reflected by the host body. As the aspect angle changes to smaller values, the ring's area that is reflecting light increases, and starts to play an important role. When the system is found in a pole-on orientation, i.e., the aspect angle is close to $0^\circ$, the rings' area is maximized contributing to a large percentage of the total brightness. Figure \ref{fig:Chariklo_examples} (a) shows a comparison between the expected evolution of Chariklo's absolute magnitude when no ring system is included (in red) and with the ring system (in blue). It can be seen how, where the system is close to $90^\circ$ or an edge-on orientation, both models agree in the expected absolute magnitude (around the years 1980 and 2008, see panel c for the evolution of the aspect angle); while for the rest of the epochs, the rings are increasing enormously the brightness of the system. In the case of $\Delta m$ (figure \ref{fig:Chariklo_examples} b), the rings produce a shield effect when measuring $\Delta m$, resulting in a slightly smaller amplitude of what it is expected for a Jacobi ellipsoid shape. 

Chiron presents a similar long-term photometric behavior as Chariklo but with a more complex scenario due to its cometary activity back in the 70s \citep[e.g.,][]{Kowal1979,Tholen1988,Meech1990}. Chiron's absolute magnitude was affected not only by the rings but also by the reflected light from the comae at that time. This was modeled in \cite{Ortiz2015} that included an exponential brightness decay for the comae behavior \cite[see figure 7 in][]{Ortiz2015}. The later work analyzed several stellar occultations produced by Chiron in different years and once again the long-term photometric models helped to distinguish between the two pole orientations offered by the analysis of the stellar occultations. Later, \cite{Cikota2018} obtained new data for both $H$ and $\Delta m$ compatible with the prediction of the models in \cite{Ortiz2015}.

In the cases of Chiron and Chariklo, rings and cometary active were found prior to the fit of the long-term models, but it is clear that the absolute magnitude is highly affected by different characteristics that the body can host and long-term photometric data can show deviations from a simple hydrostatic equilibrium. This is the case for the centaur Bienor, for which its photometric behavior clearly deviates from that of a simple Jacobi ellipsoid shape \citep{Fernandez-Valenzuela2017}. Different scenarios were studied by \cite{Fernandez-Valenzuela2017}, including rings, albedo variability, and shapes out the hydrostatic equilibrium. Although the different models can not be statistically ruled out \cite[tables in][]{Fernandez-Valenzuela2017}, it is clear that Bienor possesses some peculiarity yet to be discovered. 

To date, only three small Solar System bodies have been discovered to posses rings: Chariklo \citep{Braga-Ribas2014}, Chiron \citep{Ortiz2015,Ruprecht2015}, and Haumea \citep{Ortiz2017}. The discovery of rings has opened a new brand of research in planetary science. Although some theories have been proposed for their formation, being collisions and rotational disruptions the most plausible scenarios \citep[see][and references therein]{Braga-Ribas2014,Ortiz2015,Melita2017,Sicardy2019}, it is still not clear how they form and survive, specially for Centaurs that have suffered ``recent'' planetary encounters, or how common these features are in the outer solar region. Detecting more ring systems is mandatory to answer these open questions, however their detection is extremely difficult, with the only known ground-based method to date being stellar occultations (which would require instrumental deployment in anywhere on the Earth, depending on where the shadow of the object is produced). A careful study of existing databases and analysis of long-term photometric data can point out what targets might be the most interesting in order to put our effort in specific campaigns for stellar occultation events, to have a selection of targets to be observed with the future James Webb Space Telescope, or even visited by future spacecraft missions.

\section{Conclusions}

Long-term photometric measurements can be used to obtain different physical properties of TNOs and Centaurs, when the variability is driven by the change of the aspect angle of the object. Currently, this modellization has only been performed for four Centaurs and one TNO for which pole orientation, shape and density have been obtained (with the exception of Chiron for which density was only estimated due to its comae behaviour). Density is one of the most important properties about solar system objects, yet is one of the most difficult. To date, no more than 20 objects have density values with relative small uncertainties \cite[e.g.,][]{Kiss2019}. A larger number of objects with accurate density values is crucial to understand the evolution of the trans-Neptunian region and the material from the primitive solar nebula that was incorporated into TNOs at the time of their formation, and would help interpreting the collisional history in the outer solar system \citep{Barr2016,Bierson2019}.

Other peculiarities, such as shape out of the hydrostatic equilibrium, rings or satellites can be inferred by analyzing the long-term variability of TNOs and Centaurs. Specially, rings extremely influence the absolute magnitude of the system, and therefore, exploring existing data of different objects whether in databases (as the Small Body Node within the \href{https://pds.nasa.gov}{Planetary Data System}), or current and future surveys (e.g., the \href{https://www.planetary.org/articles/lsst-evolution-small-body-science}{Vera Rubing Survey Telescope}) might result in a source of targets in which to hunt these features; for instance, for dedicated campaigns to predict stellar occultations or even spacecraft visits. Specific campaigns to find more rings around small bodies are pivotal to statistically analyse how common these features are, and therefore, to understand their formation and survival around small bodies.

\section*{Acknowledgments}
E.~F.-V. acknowledges financial support from the Florida Space Institute and the Space Research Initiative. E.~F.-V. acknowledges Dr.~Alvaro Alvarez-Candal and Dr.~Yan Fernandez for their useful comments that help to improve this manuscript.

\bibliographystyle{frontiersinSCNS_ENG_HUMS} % for Science, Engineering and Humanities and Social Sciences articles, for Humanities and Social Sciences articles please include page numbers in the in-text citations
\bibliography{bibliography}

%%% Make sure to upload the bib file along with the tex file and PDF
%%% Please see the test.bib file for some examples of references

\end{document}